\documentclass[aps,twocolumn,preprintnumbers,showpacs,showkeys,nofootinbib,
superscriptaddress  
]{revtex4}

\usepackage{epsfig}
\usepackage{amssymb,amsmath,amsfonts,amsthm,graphicx,psfrag}
\graphicspath{{./Fig/}}                 

\newcommand{\beq}{\begin{equation}}
\newcommand{\eeq}{\end{equation}}
\newcommand{\bea}{\begin{eqnarray}}
\newcommand{\eea}{\end{eqnarray}}
\newcommand{\beas}{\begin{eqnarray*}}
\newcommand{\eeas}{\end{eqnarray*}}
\newcommand{\Fig}[1]{Fig.~\ref{#1}}
\newcommand{\Tab}[1]{Table~\ref{#1}}
\newcommand{\Sec}[1]{Section~\ref{#1}}
\newcommand{\Eq}[1]{Eq.~(\ref{#1})}

\newcommand{\aleq}{\mbox{}_{\textstyle \sim}^{\textstyle < }}
\newcommand{\ageq}{\mbox{}_{\textstyle \sim}^{\textstyle > }}

\newcommand{\Nt}{N_{\tau}}
\newcommand{\Ns}{N_\sigma}
\begin{document}
\preprint{HU-EP-12/09}
\title{Two-color QCD with staggered fermions at finite temperature \\
under the influence of a magnetic field}
\author{E.-M.~Ilgenfritz}
\affiliation{Joint Institute for Nuclear Research, VBLHEP, 141980 Dubna, Russia \\
and Humboldt-Universit\"at zu Berlin, Institut f\"ur Physik, 
12489 Berlin, Germany}
\author{M. Kalinowski}
\affiliation{Humboldt-Universit\"at zu Berlin, Institut f\"ur Physik, 
12489 Berlin, Germany}
\author{M.~M\"uller-Preussker}
\affiliation{Humboldt-Universit\"at zu Berlin, Institut f\"ur Physik, 
12489 Berlin, Germany}
\author{B.~Petersson}
\affiliation{Humboldt-Universit\"at zu Berlin, Institut f\"ur Physik, 
12489 Berlin, Germany}
\author{A.~Schreiber}
\affiliation{Humboldt-Universit\"at zu Berlin, Institut f\"ur Physik, 
12489 Berlin, Germany}

\date{June 13, 2012}

\begin{abstract}
In this paper we investigate the influence of a constant external magnetic
field on the finite-temperature phase structure and the chiral properties
of a simplified lattice model for QCD. We assume an $SU(2)$ gauge symmetry 
and employ dynamical staggered fermions of identical mass without rooting, 
corresponding to $N_f=4$ flavors of identical electric charge. For fixed 
mass (given in lattice units) the critical temperature is seen to rise with 
the magnetic field strength. For three fixed $\beta$-values, selected such 
that we stay (i) within the chirally broken phase, (ii) within the transition 
region or (iii) within the chirally restored phase, we study the approach 
to the chiral limit for various values of the magnetic field. Within the 
chirally broken (confinement) phase the chiral condensate 
is found to increase monotonically with a growing magnetic field strength.
In the chiral limit the increase starts linear in agreement with a chiral 
model studied by Shushpanov and Smilga. Within the chirally restored 
(deconfinement) phase the chiral condensate tends to zero
in the chiral limit, irrespective of the strength of the magnetic field. 
\end{abstract}



\maketitle

\section{Introduction}
\label{sec:introduction}

It is commonly believed that hadronic matter at high temperature undergoes a 
phase transition into another phase, traditionally called the ``quark-gluon 
plasma''. Actually, ab initio numerical lattice simulations have shown that the 
high-temperature behavior of QCD at low baryon number density is governed by 
two interrelated crossover phenomena, namely the transition from a low 
temperature, confined regime to a high temperature deconfined regime and the 
transition from a low temperature regime with spontaneously broken chiral 
symmetry to a high temperature regime with restored chiral symmetry. For a 
recent review see \cite{DeTar:2011nm}.
The behavior at large baryon densities, on the other hand, is only predicted 
by models, because in this case the lattice action is not real-valued in real 
QCD, which makes ab initio numerical simulations impossible. There are 
special model theories, however, where there is no sign problem at finite 
baryon density. This is the case, e.g., for $SU(2)$ gauge theory with 
dynamical fermions. 

A very interesting question is how a strong external magnetic field modifies 
the properties of strong interactions at high temperature. Cosmological models 
suggest that very strong magnetic fields ($\sqrt{eB}\sim 1-2$ GeV) have been
produced at the electroweak phase transition \cite{Vachaspati:1991nm}. 
Strong external magnetic fields cannot be ignored
in noncentral heavy ion collisions at high energy, which are produced by
the electric currents of the throughgoing spectator nucleons. Estimates of 
these fields range from $\sqrt{eB} \sim 100$ MeV for collisions at RHIC 
to $\sqrt{eB}\sim 500$ MeV at the LHC \cite{Kharzeev:2007jp,Skokov:2009qp}. 

One spectacular consequence in noncentral heavy ion collisions is the so 
called chiral magnetic effect, which was proposed 
in Ref. \cite{Kharzeev:2007jp}. In the presence of a winding number 
transition in the strongly interacting (color SU(3)) gauge field, 
the magnetic field can induce a parallel electric current.  
This leads to event-by-event fluctuations of the electric charge asymmetry 
of emitted hadrons with respect to the reaction plane. Such charge fluctuations
have, in fact, been seen in the STAR experiment at 
RHIC \cite{Voloshin:2008jx,Abelev:2009uh} and in the ALICE experiment at the 
LHC \cite{Selyuzhenkov:2011xq}. Whether these charge fluctuations actually 
come from the mechanism mentioned above, is still under debate 
\cite{Wang:2009kd,Muller:2010jd,Voronyuk:2011jd}.

In the confinement region, the presence of an external constant magnetic field
is expected to enhance the chiral symmetry breaking, i.e. to lead to an 
increase of the chiral condensate. This has been predicted in 
the Nambu-Jona-Lasinio model, as well as in the chiral model. 
Quantitatively, the growth of the chiral condensate with the magnetic field 
strength differs between the two approaches: in the NJL model the increase 
is predicted to be 
quadratic \cite{Klevansky:1989vi,Ebert:1999ht,Zayakin:2008cy}, 
while in the chiral model it is predicted to be linear in the chiral limit,
for not too strong magnetic fields \cite{Shushpanov:1997sf}. 
The latter result is due to the infrared divergences
occurring order by order in the magnetic field in the limit when the pion 
mass goes to zero. In the full solution \cite{Schwinger:1951nm} the 
divergence disappears, because the magnetic field itself acts as an infrared 
regulator.

According to several model calculations for the finite temperature transition,
the presence of an external magnetic field leads to an increase of the 
transition temperature~\cite{Klimenko:1992ch,Johnson:2008vna}. 
The effect of an external electric field has been studied within holographic
studies only \cite{Johnson:2008vna}. In a two-phase treatment of the chiral 
model \cite{Agasian:2001hv,Agasian:2008tb} 
the temperature of the chiral phase transition is seen to decrease 
with increasing magnetic field, while the originally first order transition
(suggested by a bag model) ends with vanishing latent heat, becoming a 
crossover. The possibility of a different effect of a magnetic field on the 
chiral symmetry restoring transition on one hand (strongly increasing 
transition temperature) and on the deconfining transition on the other 
(moderately increasing transition temperature) has been pointed out in 
Ref. \cite{Mizher:2010zb}. In the framework of a quark-meson model 
coupled to the Polyakov loop, the common crossover without magnetic 
field is seen to split into separate phase transitions. 

In the chiral limit, above the transition one expects the chiral order 
parameter to vanish. 
In Ref. \cite{Shushpanov:1997sf} it is shown with the help of Dyson-Schwinger 
equations for zero temperature in an external constant magnetic field 
(which is parametrically much larger than the other scales in the problem) 
that a quark mass proportional to $\sqrt{|eB|}$ is dynamically generated. 
This leads to a finite chiral condensate proportional to $|eB|^{3/2}$. 
If these considerations can be applied to the high temperature phase, too, 
even in the chiral limit the phase transition would disappear in strong 
enough magnetic fields.

Recently several lattice simulations have been performed in gauge theories 
with fermions coupled to a constant external magnetic field.%
\footnote{We emphasize that this task is conceptionally different to those 
where an external {\it colour}-magnetic field acts additionally to the 
fluctuating colour gauge field, see e.g. \cite{Cea:2007yv} for a recent work.}
The inclusion of the magnetic field does not lead to a sign problem for 
dynamical fermions. The pioneering work was performed in Refs. 
\cite{Buividovich:2008wf,Buividovich:2009ih,Buividovich:2009wi,Buividovich:2010qe}.
There the simulations were done in quenched $SU(2)$ gauge theory coupled to
overlap fermions. The chiral condensate was found to increase linearly with 
the magnetic field strength in the chiral limit even in very strong magnetic 
fields $\sqrt{eB}\leq 3$ GeV, both at zero temperature and at $T=0.82T_c$ 
slightly below the critical temperature $T_c$. A later investigation, still
within quenched QCD, tends to favor an increase of the chiral condensate 
with the magnetic field strength with an effective power $1.6(2)$ 
\cite{Braguta:2010ej}.
Simulations in full QCD with two flavors using staggered fermions have been 
reported in Refs. \cite{D'Elia:2010nq,D'Elia:2011zu}. The authors of these works find, 
in agreement with the earlier quenched simulations, that the chiral condensate 
grows with the magnetic field strength, but quadratically for not too large 
magnetic fields. They attribute this to their use of a finite quark mass, 
corresponding to a pion mass $m_{\pi}\approx 200$ Mev. 
They further find that the chiral condensate increases with the magnetic 
field in the whole range of finite temperatures, thus even in the transition 
region and beyond \cite{D'Elia:2010nq}. This, in turn, leads to an increase  
of the finite temperature transition temperature with the magnetic field 
strength. An opposite conclusion was very recently presented 
in \cite{Bali:2011qj}, namely that the transition temperature decreases 
with an increasing magnetic field strength. This comes about because the 
chiral condensate, although increasing in the confined region away from 
the transition, actually decreases with the magnetic field strength in 
the transition region. 
In this investigation the authors have performed the simulations using 
an improved action with $2+1$ flavors of dynamical staggered fermions. 
They suggest that their results (differing from others) are due to the 
fact that in their calculation the pion mass is lower, and the taste 
splitting considerably smaller. Both calculations use the ``fourth 
root trick'' for a single flavor in order to represent the right number 
of (non-degenerate) flavors. 

In this article we investigate the case of $SU(2)$ gauge theory with dynamical 
fermions at finite temperature. The influence of the constant magnetic field 
should to a large extent depend on the chiral properties, as proposed in 
Ref. \cite{Shushpanov:1997sf}. These are quite similar to those of QCD, even 
if the color group is different. Furthermore, in the $SU(2)$ theory one can
extend the investigation to finite chemical potential, as well as study in 
detail the topological excitations, which should be responsible for the 
chiral magnetic effect. In this paper we still confine our interest
to the response of the chiral condensate and the finite temperature 
transition to the magnetic field. 

We simulate the theory on the lattice, using staggered fermions, without 
invoking the root of the fermion determinant. In the continuum limit this 
leads to a four-flavor theory with equal electric charge, but we avoid
the bias due to rooting. We can determine the influence of the 
unquenching by comparing with the simulations in quenched $SU(2)$ theory 
in Refs. \cite{Buividovich:2008wf,Buividovich:2009ih,
Buividovich:2009wi,Buividovich:2010qe}. We can, of course, also test the 
validity for this case of the models mentioned above.

In \Sec{sec:actionandparameters} we define the action and the order parameters. 
In \Sec{sec:setup} we describe the setup of our simulations and
how we determine the scale by calculating the 
heavy quark potential and the pion mass at zero temperature. 
\Sec{sec:results} presents the results at finite temperature 
and with magnetic field for the chiral condensate, the Polyakov loop 
and the mean values of plaquettes differently oriented with respect to 
the magnetic field.
\Sec{sec:conclusions} contains the conclusions.

\section{Specification of the action and definition of the order parameters}
\label{sec:actionandparameters}

To describe the finite temperature theory we introduce a lattice of size 
$\Nt\times \Ns^3$.  The sites are 
enumerated by $n=(n_1,n_2,n_3,n_4)$, where the $n_i$ are integers, 
$n_i =1,2, \ldots ,\Ns$ for $i=1,2,3$ and $n_4 = 1,2, \ldots , \Nt$.
The fourth direction is the (inverse) temperature or (Euclidean) time 
direction. The lattice spacing is denoted by $a$. On the links 
$n \to n+\hat{\mu}$ we define group elements $U_{\mu}(n)\in SU(2)$, 
where $\mu = 1,2,3,4$. 
For the gauge action we adopt the usual Wilson plaquette action
\beq
S_G= \frac{\beta}{2}\sum_n \sum_{\mu<\nu} \mathrm{tr}\left(1-U_{\mu\nu}(n)\right),
\eeq
where $U_{\mu\nu}(n)$ is the $\mu\nu$ plaquette matrix attached to 
the site $n$:
\beq
U_{\mu\nu}(n) = 
U_{\mu}(n)U_{\nu}(n+\hat{\mu})U^{\dagger}_{\mu}(n+\hat{\nu})U^{\dagger}_{\nu}(n).
\eeq 
We want to introduce an external constant magnetic field
to interact with the fermions. We therefore introduce electromagnetic 
potentials in the fermion action by new, commuting group elements on the
links, namely $V_{\mu}(n)= e^{i\theta_{\mu}(n)}\in U(1)$,
with compact link angles $0 \le \theta_{\mu}(n) < 2\pi$. We further introduce
staggered fermions as Grassmann variables $\bar{\psi}(n)$ and $\psi(n)$, 
which are color vectors in the fundamental representation of $SU(2)$.
The fermionic part of the action becomes
\bea
S_F & = & a^3\sum_{p,q} \bar{\psi}(p)[D(p,q)+ ma\delta_{p,q}]\psi(q), \\
D(p,q) & = & \frac{1}{2}\sum_{\mu}\eta_{\mu}(p)[V_{\mu}(p)U_{\mu}(p)\delta_{p+\mu,q}-  
\nonumber \\
 & - & V_{\mu}^*(p-\mu)U_{\mu}^{\dagger}(p-\mu)\delta_{p-\mu,q}].
\eea
The arguments $p,q$ are 
integer four vectors denoting sites on the lattice and $\eta_{\mu}(p)$
are the usual staggered sign factors, 
\bea
& &\eta_1(n) = 1, \\
& &\eta_{\mu}(n) = (-1)^{\sum_{\nu=1}^{\mu-1}n_{\nu}}, \hspace{0.5cm} \mu=2,3,4.
\eea
In Ref. \cite{AlHashimi:2008hr} the authors propose a construction 
to include a constant magnetic field both in the continuum and on the lattice,
with periodic boundary conditions in the spatial directions. In the continuum,
to have a constant magnetic field ${\vec B}=(0,0,B)$ pointing in the 
$z$-direction they define the vector potentials as
\beq
A_{\mu}(x,y,z,t) = \frac{B}{2}(x\delta_{\mu,2}-y\delta_{\mu,1}).
\eeq
Because of the periodic boundary conditions, delta functions are needed at the 
boundary. Then ${\vec B}=\mathbf{rot} \, {\vec A}$ is constant except on the 
boundary, where we obtain a large magnetic field, so that the average of the 
magnetic field is zero
\beq
\frac{1}{L_x L_y}\int_{(x,y)-\mathrm{plane}}\, dxdy~B_z = 0.
\eeq
The translation of this construction to the lattice is a good choice,
because the lattice sees the delta function only modulo $2\pi$. 
A plaquette angle can be defined by
\beq
\theta_{\mu\nu}(n)=\Delta_{\mu} \theta_{\nu}- \Delta_{\nu}\theta_{\mu},
\eeq
where $\Delta_{\nu}f(n) \equiv (f(n+\nu)-f(n))/a$ is the lattice forward
derivative acting on 
$\theta_{\nu}(n) \sim A_{\nu}(n)$.
Then the electromagnetic plaquette part can be split as \cite{DeGrand:1980eq}
\beq
\theta_{\mu\nu} = [\theta]_{\mu\nu}(n) + 2\pi n_{\mu\nu}(n)
\eeq
with $n_{\mu\nu}(n) = -2,-1,0,1,2$. 
Plaquettes with $n_{\mu\nu}\neq 0$ are called Dirac plaquettes, 
and the reduced plaquette angle $[\theta]_{\mu\nu} \in [0,2\pi[$ corresponds
to the (gauge-invariant) electromagnetic flux through the plaquette.

A constant magnetic background field in the $z$-direction 
penetrating all the $(x,y)$ -planes of finite size $\Ns \times \Ns$
with a constant magnetic flux $\phi$ per each $(x,y)$-plaquette
is realized by the following choice:
\bea
V_1(n) =  e^{-i\phi n_2/2} \hspace{0.4cm} (n_1 =1,2,\ldots , \Ns-1), \\
V_2(n) =  e^{i\phi n_1/2} \hspace{0.5cm} (n_2=1,2,\ldots ,\Ns-1), \\
V_1(\Ns,n_2,n_3,n_4)  =  e^{-i\phi(\Ns+1) n_2/2},  \\
V_2(n_1,\Ns,n_3,n_4)  =  e^{i\phi(\Ns+1) n_1/2}.  
\eea
In order to have a constant flux also on the boundary 
it has to be quantized as follows 
($q$ denoting the same electric charge of all fermion flavors).
\beq
\phi \equiv a^2qB= \frac{2\pi N_b }{\Ns^2}, \hspace{1cm} N_b \in {\bf Z}\,. 
\label{eq:flux}
\eeq
Thus, on the lattice there is always a minimal non-vanishing flux. 
Because the angle is periodic, there is also a maximum flux. 
In fact one has to restrict oneself to
$\phi \leq \pi$, or $N_b \leq \Ns^2/2$.
At finite temperature this means that 
$\frac{\sqrt{qB}}{T}$ is restricted to the region 
\beq
\sqrt{2\pi} \frac{\Nt}{\Ns} \leq \frac{\sqrt{qB}}{T} \leq \sqrt{\pi}\Nt. 
\label{eq:limits}
\eeq
In the case of color group $SU(2)$ there is a larger chiral symmetry
than in color $SU(N)$ with $N>2$,
because the fundamental representation is equivalent to the conjugate one.
In the continuum, for $B=m=0$ there is a $U(2N_f)$ chiral symmetry, 
which is broken down to $Sp(2N_f)$ when the mass is different from zero.
If the symmetry is spontaneously broken, there are $N_f(2N_f-1)-1$ Goldstone 
bosons, where the $-1$ (one less Goldstone boson) is due to the axial anomaly. 

On the lattice, for $SU(N)$, $N>2$ there is a global $U(1)\otimes U(1)$ 
chiral symmetry for $B=m=0$. The nonsinglet axial symmetry
is broken when $m \neq 0$. Lattice calculations show that it is also
spontaneously broken, with one Goldstone boson. In color $SU(2)$ for 
$B=m=0$ there is instead a global $U(2)$ chiral symmetry on the lattice. 
For $m \neq 0, B=0$ it is broken down to $U(1)$. In the dynamical theory 
it is spontaneously broken giving rise to three Goldstone bosons. One of 
them is an electrically neutral meson, while the other two are a baryon 
and an antibaryon, which are electrically charged. Notice that in color 
$SU(2)$ the baryons are bosons. When $B \neq 0$ there is only one Goldstone 
meson. The baryons get masses proportional to $\sqrt{qB}$. The situation, 
as far as the Goldstone bosons are concerned, is thus quite similar to 
two-flavor QCD.  

In order to study the influence of the external static magnetic field on 
the chiral condensate and on the phase structure we measure the following 
(approximate) order parameters. The chiral condensate, 
which is an exact order parameter in the limit of vanishing quark mass, is 
given by
\bea
a^3<\bar{\psi}\psi>  =  - \frac{1}{\Nt\Ns^3}~\frac{1}{4}
~\frac{\partial}{\partial(ma)} \log (Z) = \nonumber \\
 =  \frac{1}{\Nt\Ns^3}~\frac{1}{4}~<\mathrm{Tr}(D+ma)^{-1}>,
\eea
where the partition function is
\beq
Z=\int \prod_{n,\mu}(d\bar{\psi}(n)d\psi(n) dU_{\mu}(n))e^{-S_G - S_F}.
\eeq
The factor $1/4$ is inserted because we define $<\bar{\psi}\psi>$ per flavor,
and our theory has $4$ flavors.

In order to locate the phase transition we use the disconnected part of the
susceptibility (later on called ``chiral susceptibility'' for simplicity),
\bea
 & \chi & = \frac{1}{\Nt\Ns^3}~\frac{1}{4}~\frac{\partial^2}{(\partial(ma))^2} 
\log (Z) \\
 & & = \chi_{\mathrm{conn}}+\chi_{\mathrm{disc}}\,, \nonumber \\
 & \chi_{\mathrm{disc}} & =  \frac{1}{\Nt\Ns^3}~\frac{1}{16}  
                  ( <(\mathrm{Tr}(D+ma)^{-1})^2> \nonumber \\
 & &  - <\mathrm{Tr}(D+ma)^{-1}>^2 ).
\eea
It is important to notice that these are bare quantities, 
which should be renormalized when comparing with continuum 
expectation values.

We further measure the average value of
the order parameter for confinement, the Polyakov loop
\bea
 &  <L> = & \frac{1}{\Ns^3}\sum_{n_1,n_2,n_3} \frac{1}{2} \times \\ 
 & &  < \mathrm{Tr} \left( \prod_{n_4=1}^{\Nt}
U_4(n_1,n_2,n_3,n_4)\right) > \nonumber
\eea
and the corresponding susceptibility
\beq
\chi_L = \Ns^3(<L^2> - <L>^2).
\eeq

Intuitively it is clear that a constant magnetic field oriented
into one of the three space directions has to violate the 3d isotropy 
of the system. The isotropy violation can only be caused by the coupling
of the magnetic field to the fermionic part of the action and should become
visible within the effective gauge action after the fermion degrees have been 
integrated out. 
In order to demonstrate this anisotropy one can easily compute the 
average non-Abelian plaquette values separately for the different 
space-time planes $(\mu,\nu)$ (an averaging over the lattice is implied)
\beq
P_{\mu\nu}  =  <\frac{1}{2}~\mathrm{Re}~\mathrm{Tr}~U_{\mu\nu}>\,, \\
\label{eq:anisotropy}
\eeq
as a function of the magnetic field strength.

\section{Setup of the simulations} 
\label{sec:setup}
In order to simulate the model with dynamical staggered fermions ($N_f=4$)
we employed the standard Hybrid Monte Carlo algorithm. We chose the number of 
integration steps $n_{int}$ and their ''time'' length $\delta \tau$ 
such that the length of a trajectory was $n_{int} \delta\tau=1$ while 
the acceptance rate was larger than $0.8$. For the finite-temperature 
measurements we used a $16^3 \times 6$ lattice. We measured the chiral 
condensate on every fifth configuration, the Polyakov loop and the 
plaquette variables on every configuration. 
The chiral condensate was calculated with the random source method. 
Thereby we used 100 random sources per configuration.
The integrated autocorrelation time was taken into account in all our 
error estimates. In general, apart from simulating very near to the 
transition temperature, the integrated autocorrelation times of all observables 
were estimated mostly well below 20. The number of configurations 
(trajectories) generated within a run varied between 1800 and 5000. 
In general, 300 configurations were discarded for initial thermalization. 

We also made a scan of the susceptibilities $\chi_{disc}$ and 
$\chi_L$ near the transitions for different magnetic fields. 
For this aim we generated longer runs of length between $10^4$ 
and $2 \cdot 10^4$ trajectories, while we measured the chiral 
condensate and the Polyakov loop with the usual frequency.
The autocorrelation times were estimated to $O(25)$ near the 
transition points. 

Some zero-temperature simulations 
on a lattice of size $16^3\times 32$ without magnetic field
were performed additionally in order to estimate the
lattice spacing, the pion mass, and the critical temperature 
in physical units. 
We are, of course, aware of the fact that we are considering 
a fictitious world of two-color QCD with four flavors
of identical electric charge $q$.
Nevertheless, a scale determination provides a rough estimate 
how near we are to the chiral limit, and how large 
the explored magnetic field strengths are.
The number of trajectories per $\beta$-value 
used in the $T=0$ case was 2000. Measurements were performed 
after every fourth trajectory. 

At $T=0$ we first calculated the pion propagator $C_{\pi}(t)$ 
for the three $\beta$-values 1.7, 1.8 and 1.9, fitting it to the 
usual form ($N_\tau = 32$)
\beq
C_{\pi}(t) = C e^{-E t} + C e^{E (t-N_\tau)}.  
\label{eq:pion}
\eeq 
The fit parameters in lattice units can be found in \Tab{tab:tab1}.
Our results are compatible with earlier ones obtained for 
larger quark masses on smaller lattices \cite{Laermann:1986pp}.

Moreover, we computed the heavy quark potential 
from HYP-smeared~\cite{Hasenfratz:2001hp} Wilson loops 
(with HYP-smearing in the version of Ref.~\cite{Bornyakov:2005iy}) 
and for comparison also with APE-smearing~\cite{Albanese:1987ds} 
at the three $\beta$-values 1.8 , 1.9 and 2.1, by fitting it
to the form
\beq
V(\vec{R}) = V_S(R) + C(\frac{1}{R}-G_{L}(\vec{R}))
\eeq
with
\beq
V_S(R) = A - B/R + \sigma R,
\label{eq:potential}
\eeq
where $G_{L}(\vec{R})$ is the free gluon propagator on the lattice. 
This correction is only important at short distances. For details 
of this procedure see \cite{Schilling:1993bk}. 
The fit results obtained from the HYP-smeared data  
are presented in \Tab{tab:tab1}, too.
\begin{table*}
\mbox{
\setlength{\tabcolsep}{1.0pt}
\begin{tabular}{|c||c|c|c|c||c|c|c|c|}
\hline
 & \multicolumn{4}{|c||}{$C_{\pi}(t)$}
 & \multicolumn{4}{|c|}{$V_S(R)$} \\
\hline
 $\beta$ & $t_{min}$ &$C$ & $E=am_\pi$ &  $\chi^2_{dof}$ 
                                       & $A$ & $B$ & $\sigma$ & $\chi^2_{dof}$ \\
\hline
  1.7 & 5 &1.48(5)   & 0.265(1) & 0.047   & - & - & - & - \\
  1.8 & 6 &1.01(3)   & 0.285(1) & 0.023   & 0.265(38) & 0.370(42) &  0.169(8)   &  0.872  \\
  1.9 & 8 &0.311(11) & 0.296(2) & 0.039   & 0.265(9)  & 0.300(11) &  0.0725(19) &  0.681  \\
  2.1 & - & -        &  -       & -       & 0.250(9)  & 0.265(13) &  0.0170(16) &  1.93  \\
\hline
\end{tabular}
}
\caption{Fit parameters in lattice units for the pion correlator $C_{\pi}(t)$ acc. 
to \Eq{eq:pion} and for the static potential $V_S(R)$ acc. to \Eq{eq:potential}. 
The fit range for the pion correlator starts at lattice distance $t_{min}$.}
\label{tab:tab1}
\end{table*}

\begin{figure*}[h!]
\includegraphics[width=1.0\textwidth]{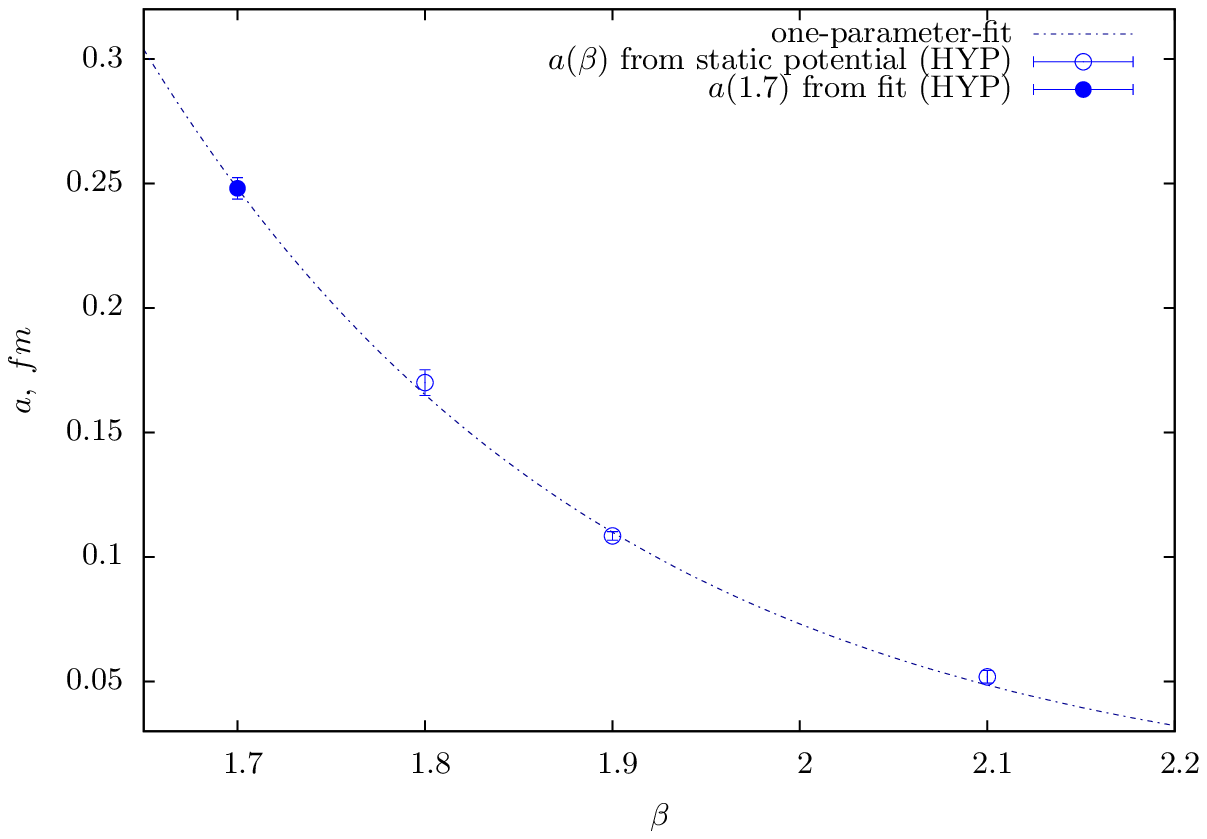}
\vspace*{-15cm}
\caption{The lattice spacing $a$ vs. $\beta$. The two-loop beta-function 
acc. to \Eq{eq:scaling} was used as fit-function, and  $a(\beta =1.7)$ was 
obtained through extrapolation.}
\label{fig:fig1}
\end{figure*}

In order to estimate the lattice spacing in physical units we have determined 
the Sommer scale parameter $R_0$ \cite{Sommer:1993ce} related to the static force 
$~F_S(R) \equiv dV_S/dR~$ through the condition 
\beq
F_S(R_0) R_0^2 = 1.65\,. 
\label{eq:sommerscale}
\eeq
The values $R_0/a$ found at $ma=0.01$ for the three $\beta$-values (1.8, 1.9, 2.1) 
are given in \Tab{tab:tab2}. To get the scale in 
physical units, we adopted the physical value $r_0=0.468(4)$ fm
fixed for QCD \cite{Bazavov:2011nk}. Our measurements of the scale $a$ at 
the three $\beta$-values are well fitted by the two-loop formula for the 
$\beta$-function for the $SU(2)$ case and four flavors 
\bea \label{eq:scaling}
a(\beta) = \frac{1}{\Lambda_L} (\frac{4\beta_{0}}{\beta})^{\frac{-\beta_{1}}{2\beta_{0}^2}}
~\exp(\frac{-\beta}{8\beta_{0}}),  \\
\beta_{0}= \frac{7}{24}\pi^{-2},  \qquad \quad \beta_{1}= \frac{19}{384}\pi^{-4} 
\nonumber
\eea
with $\Lambda_L=0.00660(12)~\mathrm{fm}^{-1}$ and $\chi^2_{dof}=1.64$ 
(see \Fig{fig:fig1}).
Thus, we used this formula for an extrapolation down to $\beta=1.7$, where 
the lattice spacing $a$ is too large to allow a safe determination of $R_0$. 
In this way we could roughly estimate the pion mass and the magnetic 
field strength also at this value of the bare coupling and $ma=0.01$.
In \Tab{tab:tab2} we have collected all our results for setting the scale.
 
Let us mention, that for finite temperature with $\Nt=6$ and $ma=0.01$ 
(for vanishing magnetic field strength) the critical value $\beta_c$ will be seen 
close to $\beta=1.8$. Thus, we are able to estimate also the critical temperature 
in physical units. We find $T_c \simeq 193(6)$ MeV.
\begin{table*}
\mbox{
 \begin{tabular}{|c|c|c|c|c|}
\hline
 $\beta$ & $R_0 / a$ & $a [fm]$ & $m_\pi[MeV]$ & $\sqrt{qB}_{N_{b}=50}[GeV]$ \\
 \hline
  1.7 & 1.89(4)  & 0.248(4)  & 210(4)   & 0.881(15) \\ 
  1.8 & 2.75(8)  & 0.170(5)  & 330(10)  & 1.29(4)  \\ 
  1.9 & 4.32(6)  & 0.108(2)  & 537(9)   & 2.02(3)  \\ 
  2.1 & 9.03(43) & 0.052(2)  & -        & 4.22(21)  \\  
\hline
\end{tabular}
}
\caption{Results in physical units for the Sommer scale $R_0$, the 
lattice spacing $a$, the pion mass $m_\pi$, and the quantity $\sqrt{qB}$ 
characterizing the magnetic field strength for $N_{b}=50$ flux units for 
various $\beta$-values considered in \Sec{sec:results}.  
The values for $\beta = 1.7$ were estimated by extrapolating with the 
two-loop beta-function.}
\label{tab:tab2}
\end{table*}

\section{Results at finite temperature}
\label{sec:results}

\begin{figure*}[tb]
\includegraphics[width=1.0\textwidth]{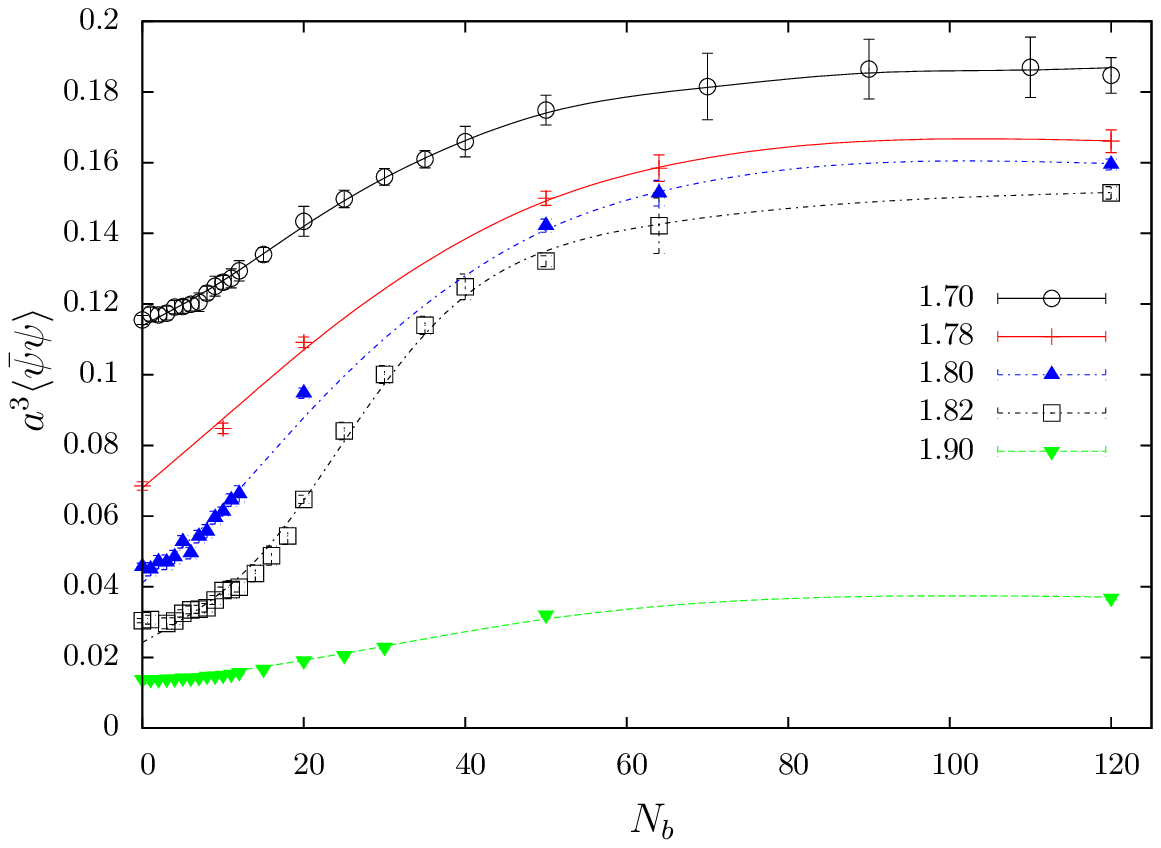}
\vspace*{-15cm}
\caption{The rising and saturation behavior of the chiral condensate as a 
function of the magnetic field in flux units for $am=0.01$ and for various 
$\beta$-values.  The lattice size is $16^3 \times 6$. 
The curves are to guide the eyes. 
}
\label{fig:fig2}  
\end{figure*}

\begin{figure*}[tb]
\includegraphics[width=1.0\textwidth]{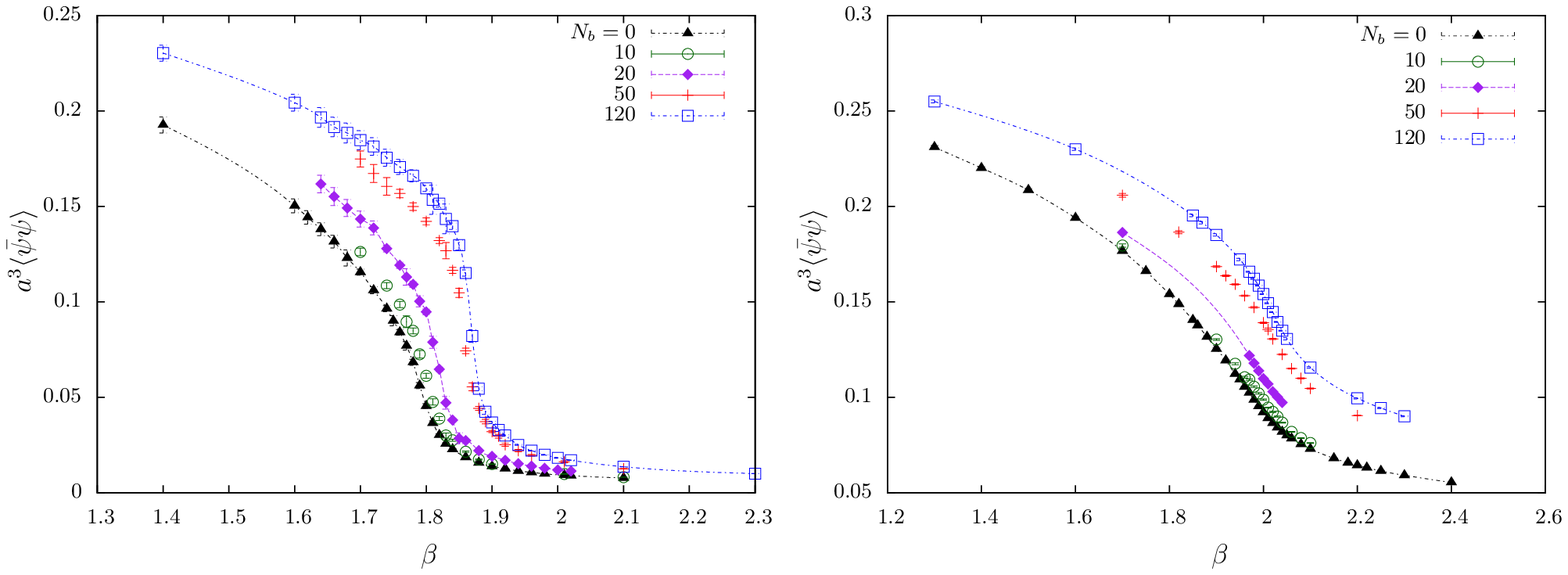}
\vspace*{-15cm}
\caption{The bare chiral condensate $a^3 \langle\bar{\psi}\psi\rangle$ vs. 
inverse coupling $\beta$ for various magnetic fluxes $\phi$ (in flux units) 
and for two different bare masses $ma=0.01$ (left panel) and 
$am=0.1$ (right panel). The lattice size is $16^3 \times 6$.
The curves are to guide the eyes. 
}
\label{fig:fig3}
\end{figure*}

The finite temperature simulations are performed on lattices of 
size $16^3\times 6$.
In \Fig{fig:fig2} the bare chiral condensate is shown as a
function of the flux number $N_b$ for various values
of $\beta$ in the region of the transition value $\beta=\beta_c=1.8$.
Because the lattice is finite the chiral condensate is 
periodic in the magnetic flux $\phi$.
We see that saturation effects set in at $N_b\approx 60$. 
This corresponds to $N_b\approx \Ns^2/4$ or $\phi\approx \pi/2$,
which is half of the maximally achievable flux, see formula (\ref{eq:limits}). 
Our investigation is mostly restricted to values of $N_b\leq 50$. 
The corresponding maximal values of $\sqrt{qB}$ are given in \Tab{tab:tab2}. 
They are in general
greater than the physical region of interest, $\sqrt{qB}\aleq 1$ GeV.
In \Fig{fig:fig3} we plot the bare chiral condensate as a function 
of $\beta$ for a set of numbers of flux quanta for the bare quark 
mass $ma=0.01$ and $0.1$, respectively. For the smaller quark mass we see quite
clearly a transition for all values of the flux quanta
$N_b$ under consideration.
It is important to notice that the bare chiral condensate increases 
with increasing magnetic field for fixed $\beta$, for all values of 
$\beta$ in the transition region. This indicates that the chiral 
transition moves to higher temperatures as the magnetic field is 
increasing. This tendency is in agreement with the results in
\cite{D'Elia:2010nq,D'Elia:2011zu} but opposite to the tendency seen
in \cite{Bali:2011qj} where the bare chiral condensate decreases 
with the flux $\phi$ in the transition region, leading to a decrease of
the transition temperature. As shown in Section III, 
without magnetic field, for $am=0.01$ and 
close to $T_c$ we have reached a ratio $m_{\pi}/T_c \approx 1.7$, 
which is similar to the ratio in 
\cite{D'Elia:2010nq,D'Elia:2011zu}, but higher than that in 
\cite{Bali:2011qj}.  We also have a different gauge group.
It is clear that our observation does not represent a direct contradiction
to the results in \cite{Bali:2011qj}. 

For the higher quark mass $ma=0.1$ the transition in the chiral condensate 
turns out very smoothly.

\begin{figure*}[tb]
\includegraphics[width=1.0\textwidth]{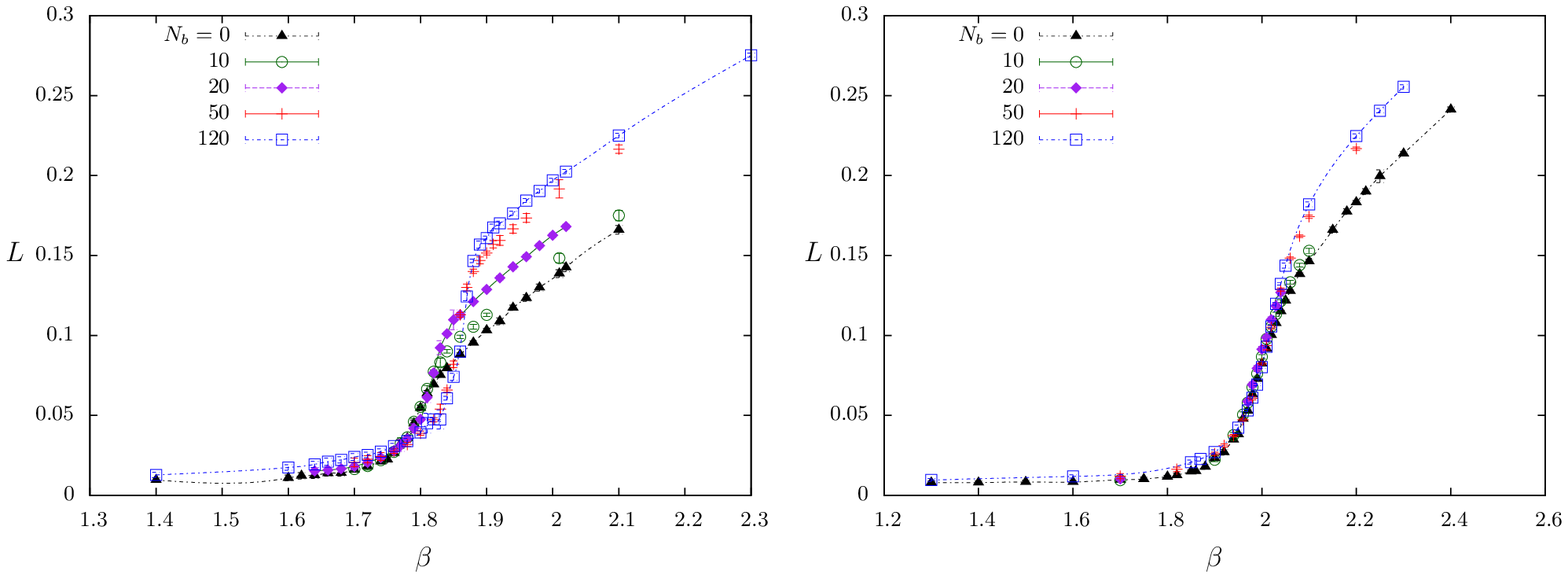}
\vspace*{-15cm}
\caption{The (unrenormalized) Polyakov loop expectation value $<L>$ vs.
inverse coupling $\beta$ for various magnetic fluxes $\phi$ and for two 
different bare masses $ma=0.01$ (left panel) and $am=0.1$ (right panel). 
The lattice size is $16^3 \times 6$.
The curves are to guide the eyes. 
}
\label{fig:fig4}
\end{figure*}

In \Fig{fig:fig4} the expectation value of the Polyakov loop is shown 
vs. $\beta$ for the same two values of the bare quark mass.
This is an indicator for the deconfinement transition. For the lower quark mass
the deconfinement transition and the chiral transition are consistent with
happening at the same temperature.
Furthermore, the transition temperature increases with the quark mass.
At the high quark mass there seems to be only a weak effect of the magnetic
field on the deconfinement temperature.

\begin{figure*}[tb]
\includegraphics[width=1.0\textwidth]{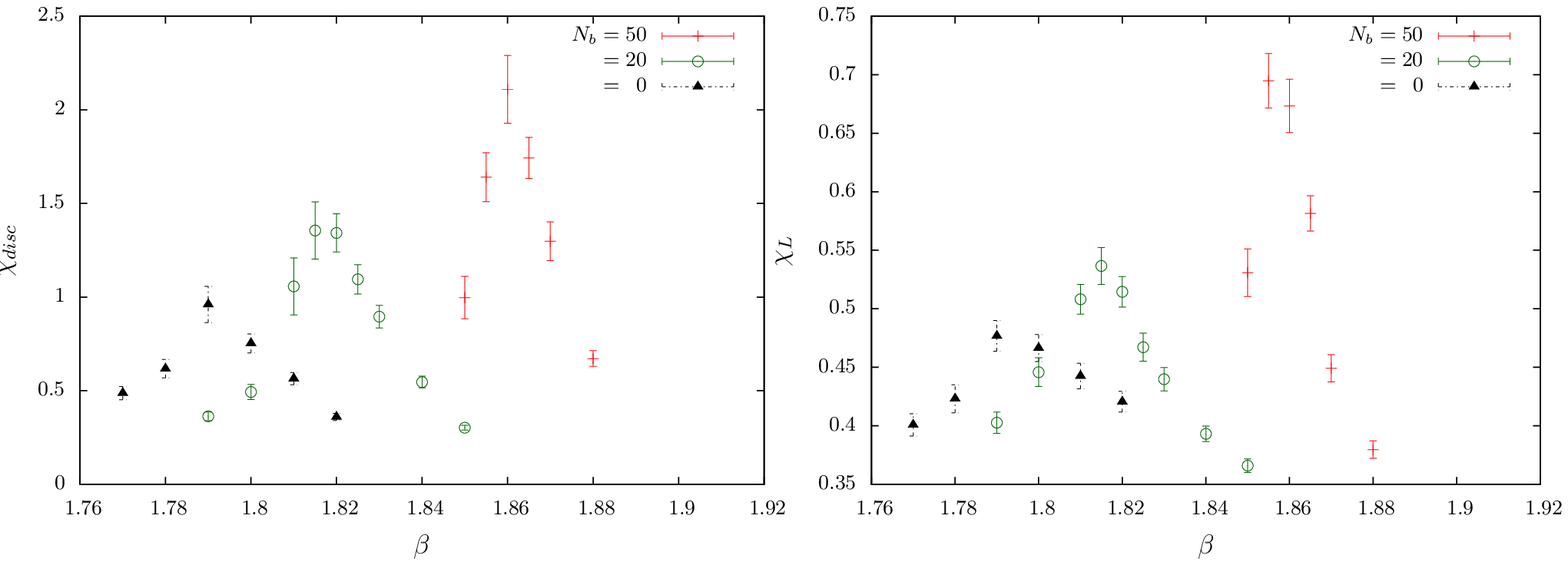}
\vspace*{-15cm}
\caption{The chiral susceptibility $\chi_{disc}$ (left panel) 
and the Polyakov loop susceptibility $\chi_{L}$ (right panel), 
shown versus $\beta$ at $am=0.01$ for various magnetic 
fluxes corresponding to the lattice size $16^3 \times 6$ (see text). 
For the computation of $\chi_{disc}$ 100 stochastic sources per 
configuration have been used.}
\label{fig:fig5}
\end{figure*}

To study the change of the transition temperature
at the lower quark mass more quantitatively we have
calculated the susceptibilities.
In \Fig{fig:fig5} we show the chiral susceptibility and the Polyakov loop 
susceptibility for the low quark mass $ma=0.01$.
It is clearly seen in the left figure that the 
chiral transition indeed moves to higher temperatures
as the magnetic field becomes stronger. In the right figure we show
the same effect for the Polyakov loop
susceptibility. In fact, the maxima of the two susceptibilities are at the
same value for given magnetic field. 
There is no sign of a splitting between the chiral and the deconfinement 
transition. The height of the peaks increases with the field
strength. However, only a finite size scaling analysis could show if the 
transition remains a crossover or becomes a real phase transition. 

In order to study the dependence of the chiral condensate on the magnetic 
field strength in the chiral limit, we have chosen to investigate in more 
detail the behavior of the chiral condensate as a function of the 
quark mass and the magnetic field for three fixed values of $\beta$.
Because we keep $\beta$ fixed as we vary the quark mass and the magnetic 
field, the lattice spacing $a$ is also fixed and we have eliminated
lattice effects coming from the variation of $a$.
The three values are $\beta = 1.70$ (which is clearly in the confined phase),
$\beta=1.90$ (which is in the transition region), and $\beta=2.10$ (which is 
in the deconfined phase). 
Let us first discuss the results in the confined region.

\begin{figure*}[tb]
\includegraphics[width=1.0\textwidth]{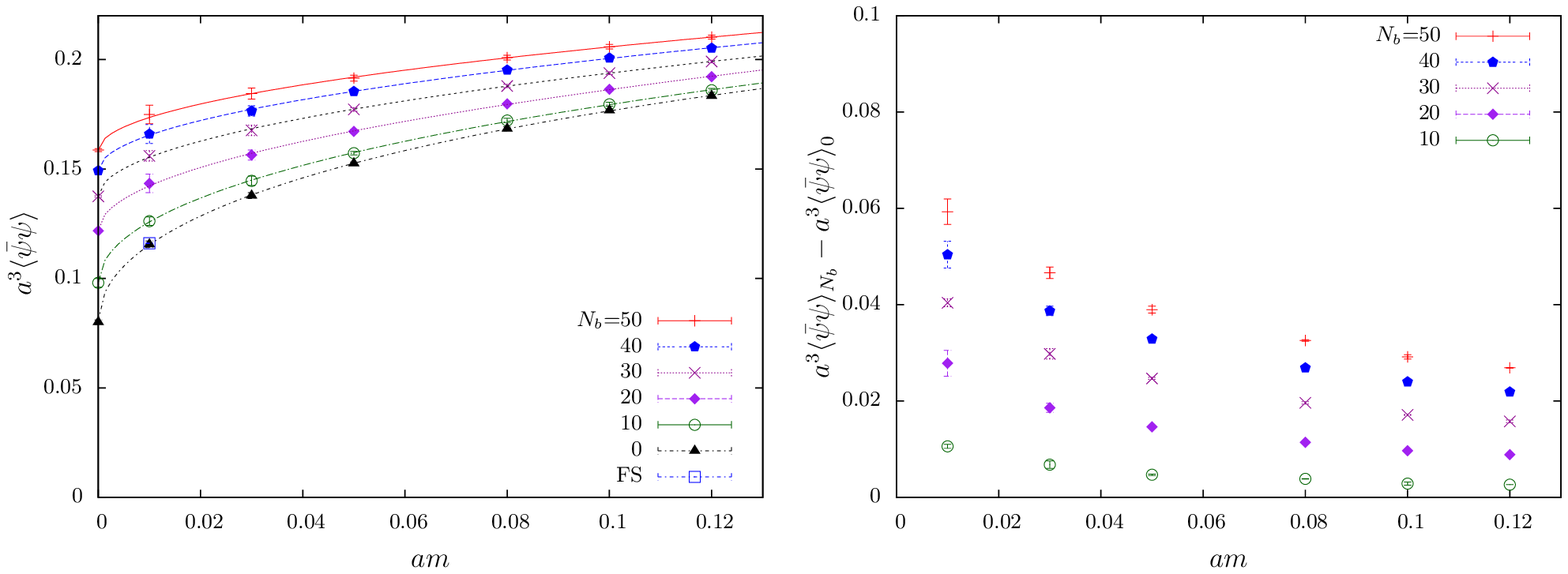}
\vspace*{-15cm}
\caption{Mass dependence of the bare chiral condensate (left) and 
of the subtracted chiral condensate (right) for various magnetic fluxes 
within the confinement phase ($\beta=1.70$). The lattice size is 
$16^3 \times 6$. In order to check for the smallness of finite-size 
effects even for the lowest quark mass we show also a data 
point (FS) obtained with $24^3 \times 6$. The lines correspond to fits
with the chiral fit function $f_1(ma)$ according to \Tab{tab:tab3}.}
\label{fig:fig6}
\end{figure*}

\begin{figure*}[tb]
\includegraphics[width=1.0\textwidth]{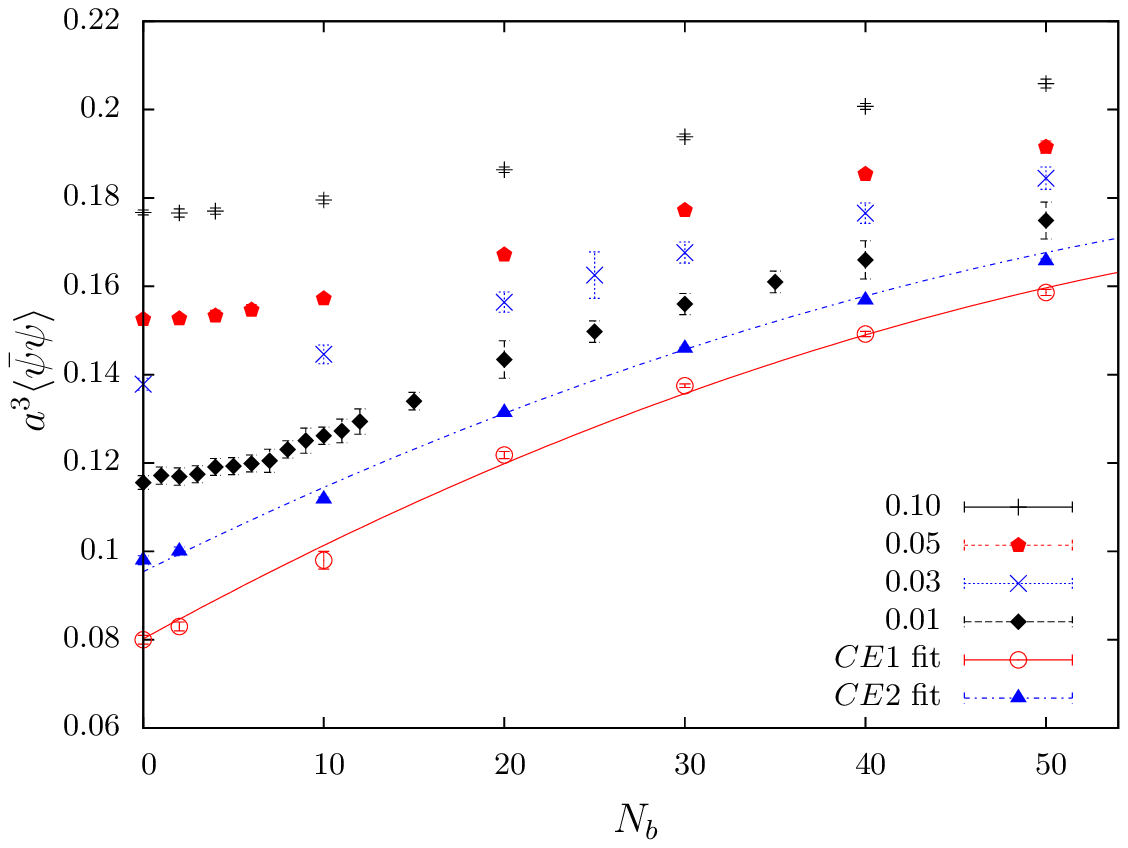}
\vspace*{-15cm}
\caption{The bare chiral condensate as a function of the flux for different 
masses and extrapolated to the chiral limit with $f_1$ acc. to \Eq{eq:3d} 
(circles) and $f_2$ acc. to \Eq{eq:4d} (triangles), respectively. The dotted 
lines show the corresponding chiral limit fit results CE1 and CE2 with the 
polynomial ansatz \Eq{eq:chirallimit}. }
\label{fig:fig7}
\end{figure*}

In the left panel of \Fig{fig:fig6} we show the dependence of the bare chiral 
condensate on the quark mass for various values of the magnetic flux. 
To obtain the results relevant to continuum physics, one has to subtract 
an additive divergence for finite quark mass, as well as do a multiplicative 
renormalization, which is needed also at zero mass.
In the right panel of \Fig{fig:fig6} we show the difference between the bare chiral
condensate for finite fluxes subtracted by the same quantity at zero flux. 
This eliminates the main part of the additive divergence. In the left panel
we have also included points at quark mass zero, where there are no additive 
divergencies. The values at zero quark mass are obtained by a chiral 
extrapolation. We perform this extrapolation in two ways. Because we are not
very far from the transition, we suppose that we can use the formula for the
reduced three dimensional model \cite{Pisarski:1983ms}.
See also \cite{Wallace:1975vi,Hasenfratz:1989pk,Smilga:1993in}. 
In that case we may use the ansatz $a^3<\bar{\psi}\psi>=f_1(ma)$, where
\beq
 f_1(ma) = a_0 + a_1 \sqrt{ma} + a_2 ma 
\label{eq:3d}
\eeq
with the non-analytic term coming from the Goldstone bosons. 
Such a parametrization has been used in \cite{Bazavov:2011nk} 
in the context of the finite temperature transition in QCD. 
As an alternative we also use the chiral extrapolation
relevant for zero temperature, namely
$a^3<\bar{\psi}\psi>=f_2(ma)$, where
\beq
 f_2(ma) = b_0 + b_1 ma \log ma + b_2 ma 
\label{eq:4d}.
\eeq
The fits performed with the Eq. (\ref{eq:3d}) are shown as dashed or dotted 
lines in \Fig{fig:fig7}. The fit parameters of both kinds of fits
are summarized for $\beta=1.70$ and various magnetic fluxes in \Tab{tab:tab3}. 
\begin{table*}
\mbox{ 
\setlength{\tabcolsep}{1.0pt}
\begin{tabular}{|c|c|c|c|c|c|c|c|c|} 
\hline 
 & \multicolumn{4}{|c|}{$f_1(ma)$} 
 & \multicolumn{4}{|c|}{$f_2(ma)$} \\
\hline
$N_b$ & $a_0$ & $a_1$ &  $a_2$ & $\chi^2_{dof}$ & $b_0$ & $b_1$ & $b_2$ & $\chi^2_{dof}$ \\
\hline
  0 & 0.080(1) & 0.36(1) & -0.19(2) & 0.33   & 0.098(1) & -0.42(1) & -0.20(2) & 0.73 \\
  2 & 0.083(1) & 0.34(1) & -0.16(2) & 0.20   & 0.100(1) & -0.42(1) & -0.16(3) & 0.30 \\
 10 & 0.098(2) & 0.28(1) & -0.09(3) & 0.66   & 0.1118(5)& -0.28(1) &  0.08(1) & 0.10 \\
 20 & 0.1218(8)& 0.203(3)& 0        & 0.73   & 0.1314(2)&-0.238(1) &  0       & 0.80 \\
 30 & 0.1375(4)&0.177(1) & 0        & 0.20   & 0.1459(3)&-0.208(2) &  0       & 0.18 \\
 40 & 0.1492(6)&0.162(2) & 0        & 0.35   & 0.1568(2)&-0.1901(9)&  0       & 0.05 \\
 50 & 0.1586(7)&0.149(2) & 0        & 0.23   & 0.1657(5)&-0.174(2) &  0       & 0.18 \\
\hline
\end{tabular}
}
\caption{Chiral fit parameters for the fits $f_1(ma)$ (\Eq{eq:3d}) and  $f_2(ma)$ (\Eq{eq:4d}) 
allowing to extrapolate to the chiral limit for $\beta=1.70$ (chirally broken phase) and
various magnetic field strengths. 
The fit curves obtained with $f_1$ are shown in the left panel of \Fig{fig:fig6}.}
\label{tab:tab3}
\end{table*}
\begin{table*}
\mbox{ 
\setlength{\tabcolsep}{1.0pt}
 \begin{tabular}{|c|c|c|c|c|}
\hline
    & \multicolumn{4}{|c|}{$g(N_b)$}    \\
 \hline
    & $c_0$ & $c_1$ &  $c_2$ & $ \chi^2_{dof}$  \\
 \hline
  CE1 & 0.080(1) & 0.027(2) & -0.00015(5) & 1.18 \\     
  CE2 & 0.095(2) & 0.021(2) & -0.00015(3) & 9.4  \\
\hline 
\end{tabular}
}
\caption{Fit parameters for the chiral condensate as a function of the
magnetic flux $N_b$ in the chiral limit acc. to the polynomial ansatz 
\Eq{eq:chirallimit}. 
CE1 and CE2 correspond to the chiral fit extrapolations with fit 
functions $f_1$ and $f_2$, respectively, as presented in 
\Tab{tab:tab3}. Both the fit results are shown in \Fig{fig:fig7}.}
\label{tab:tab4}
\end{table*}

The value of the bare chiral condensate in lattice units within the chiral 
limit ($m=0$) for various values of the flux are given by $a_0$ and $b_0$,
respectively. 
One can see that the statistical errors on the parameters are much smaller 
than the systematic errors coming from the fit formulas. 
We are, however, not interested in the absolute values of the bare chiral 
condensate at zero mass, but in the dependence of this value on the magnetic 
field strength. The flux dependent part does not need renormalization, 
and thus can be directly compared to continuum models. 
In \Fig{fig:fig7} we show the bare chiral condensate as a function of the
number of magnetic flux quanta for different bare quark masses. The chirally
extrapolated points are obtained with the fit functions in Eqs. (\ref{eq:3d}) 
and (\ref{eq:4d}). One can see that for finite quark mass the data at small 
magnetic flux are consistent with a quadratic behavior, while 
the extrapolated values to $ma=0$ seem to start with a linear behavior
irrespective of the specific chiral extrapolation $f_1$ or $f_2$ used.
This behavior is in agreement with the prediction of the chiral model 
of Ref. \cite{Shushpanov:1997sf}. 
In \Tab{tab:tab4} we provide the fit parameters for the chiral condensate 
as a function of the magnetic flux $\phi \sim N_b$ 
\beq
g(N_b)=c_0(1+c_1 N_b+c_2 N_b^2)
\label{eq:chirallimit}
\eeq
in the chiral limit obtained with the chiral extrapolation $f_1$ (CE1) and
$f_2$ (CE2), respectively. 
We have not calculated the pion decay constant, and therefore we do not
make a quantitative comparison with the chiral model at finite temperature
and magnetic field studied in Ref.~\cite{Agasian:2001hv}.

\begin{figure*}[tb]
\includegraphics[width=1.0\textwidth]{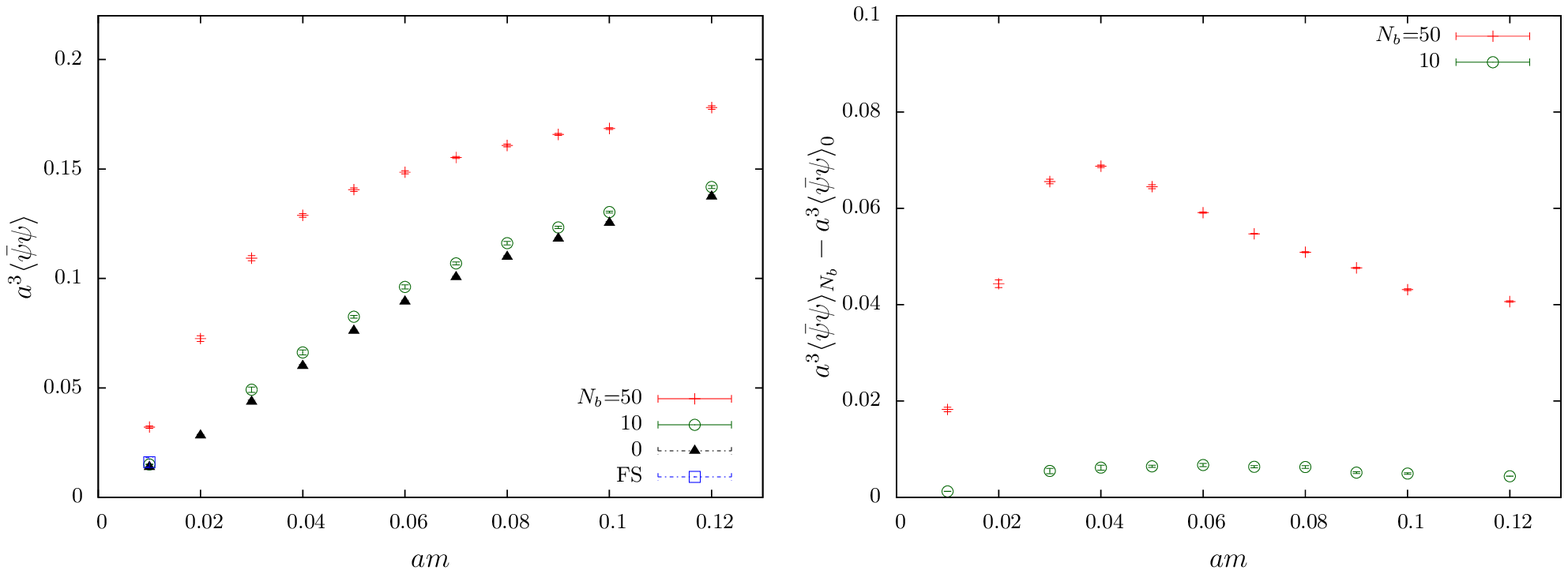}
\vspace*{-15cm}
\caption{Mass dependence of the bare chiral condensate (left) and of the 
subtracted chiral condensate (right) for three magnetic flux values within 
the transition region ($\beta=1.90$). The lattice size is $16^3 \times 6$. 
For the lowest mass we show also a data point (FS) obtained with lattice 
size $24^3 \times 6$.}
\label{fig:fig8}
\end{figure*}

\begin{figure*}[tb]
\includegraphics[width=1.0\textwidth]{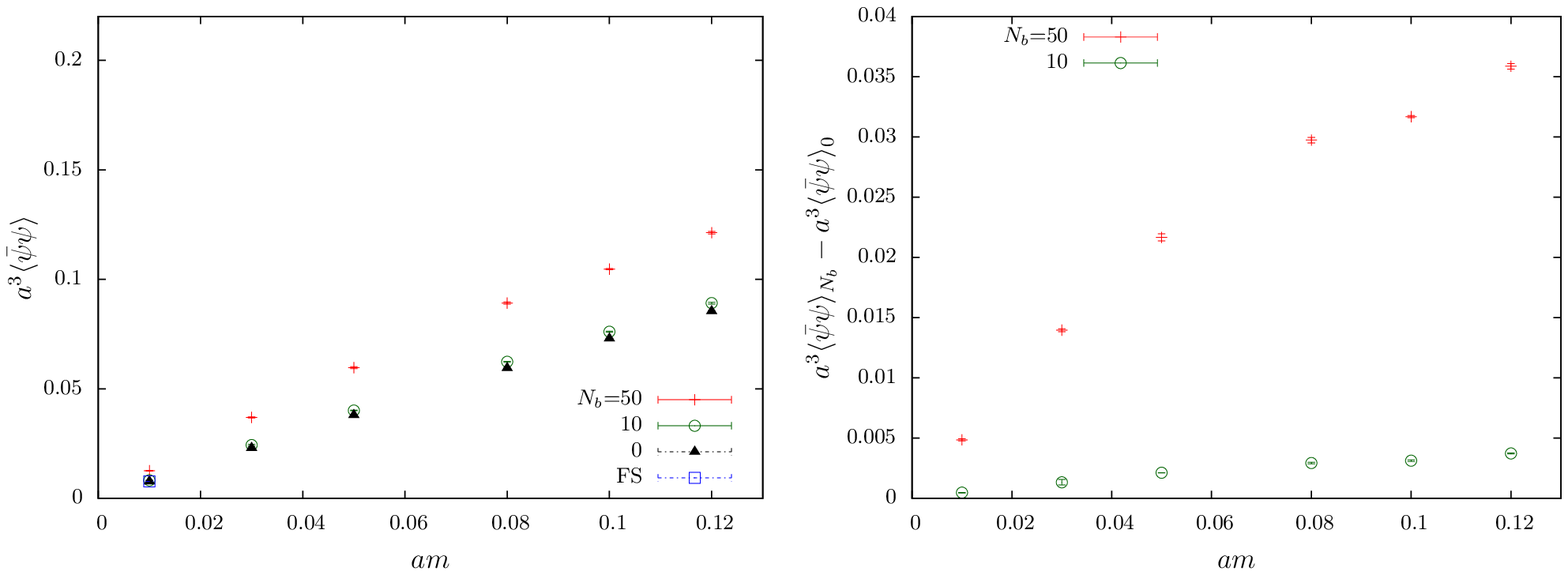}
\vspace*{-15cm}
\caption{Same as in \Fig{fig:fig8} but within the deconfinement phase 
($\beta=2.1$).} 
\label{fig:fig9}
\end{figure*}

Now we come to the transition region and the temperature region above 
the transition.
In \Fig{fig:fig8} we show the mass dependence of the bare chiral condensate
(left panel) and the subtracted chiral condensate (right panel) 
in the transition region (at $\beta=1.90$) for three values of the 
magnetic flux. One can see that for finite flux, as well as for zero flux,
the bare and subtracted chiral condensates are consistent with extrapolating 
to zero in the chiral limit. For the highest flux, $N_b=50$ one can clearly
discern two regions of behavior. For $am \ageq 0.04$ the chiral condensate 
seems to extrapolate to a finite value, but for $am \aleq 0.04$ it actually 
extrapolates to zero. This can be understood, if one assumes that the 
transition for $N_b=50$ at this value of $\beta$ takes place for 
$am\approx 0.04$. In \Fig{fig:fig9} we present the same quantities as above, 
but for $\beta=2.10$. This is well inside the chirally restored phase. 
The chiral condensate extrapolates to zero for all values of the flux.
Thus chiral symmetry is restored for all values of the flux that we have 
investigated.
Within our precision there is no evidence for a quark mass which is 
spontaneously created by the magnetic field as suggested in a selfconsistent 
calculation in \cite{Shushpanov:1997sf} using the Dyson-Schwinger equations.

Finally, we study the indirect influence of the magnetic field on the gluonic 
part of the action. For this purpose we provide the plaquette energies 
$1-P_{\mu\nu}$ individually for the six differently oriented plaquettes, 
as given in Eq.(\ref{eq:anisotropy}) for varying fermionic mass.
We do this for the same values of $\beta$ as above, namely 
$\beta=1.70, 1.90$ and $2.10$ and for two values of the flux corresponding
to $N_b=0$ (zero magnetic field) and $N_b=50$ (strong magnetic field). 

\begin{figure*}[tb]
\includegraphics[width=1.0\textwidth]{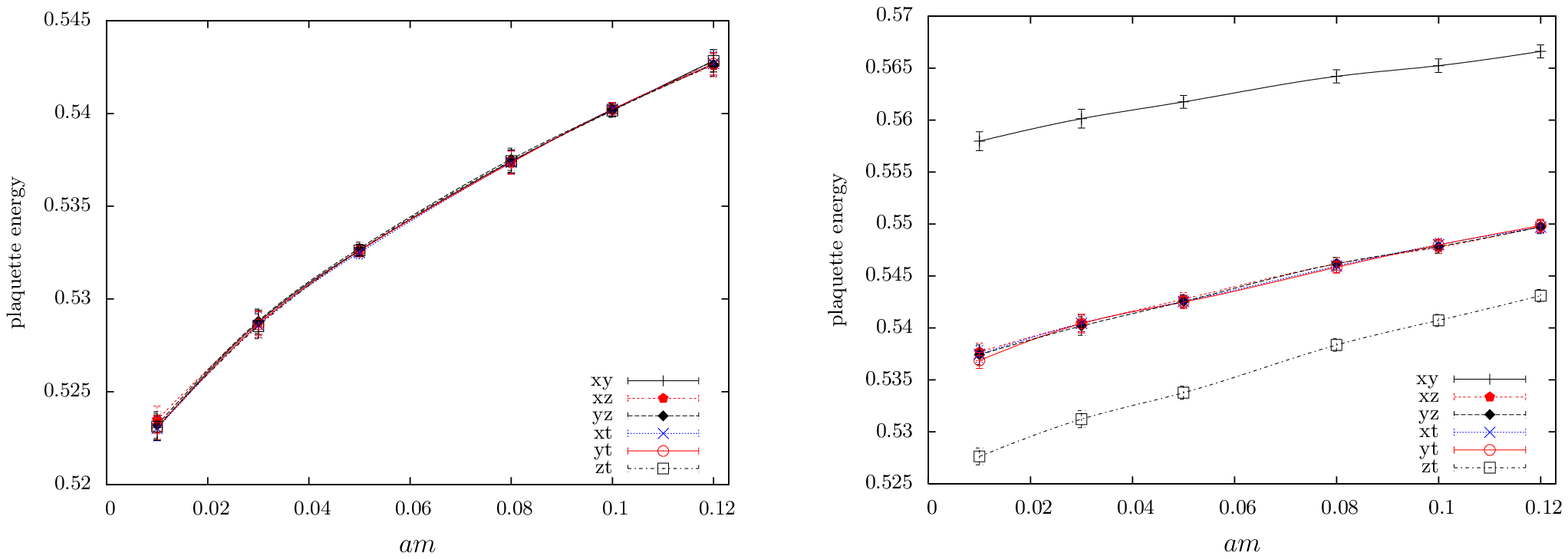}
\vspace*{-15cm}
\caption{Plaquette energies $(1-P_{\mu\nu})$ in the confinement 
phase ($\beta=1.70$) without (left) 
and with (right) an external magnetic field ($N_b=50$).
The curves are to guide the eyes. 
}
\label{fig:fig10}
\end{figure*}

In \Fig{fig:fig10} one can see that at $\beta=1.70$ (within the confined phase)
all the plaquettes are degenerate for vanishing magnetic field. For a 
strong magnetic field pointing into the $z$ direction, the $xy$ and the $zt$ 
plaquettes become well separated from the others. This effect can be qualitatively
understood by representing the effective gauge action in terms of fermionic loops
within a hopping parameter expansion for large fermion mass $am \sim 1/\kappa$. 
While in the order $\kappa^4$ of such an expansion only $xy$-plaquettes receive 
a coupling to the magnetic field strength, in the order $\kappa^6$ loops 
extending into three directions $xyz$ or $xyt$ may couple to it. 
The $zt$-plane is distinct, since only in the order $\kappa^8$ and beyond 
there exist loops extending into all four directions which couple to the 
magnetic field in the effective action.    

\begin{figure*}[tb]
\includegraphics[width=1.0\textwidth]{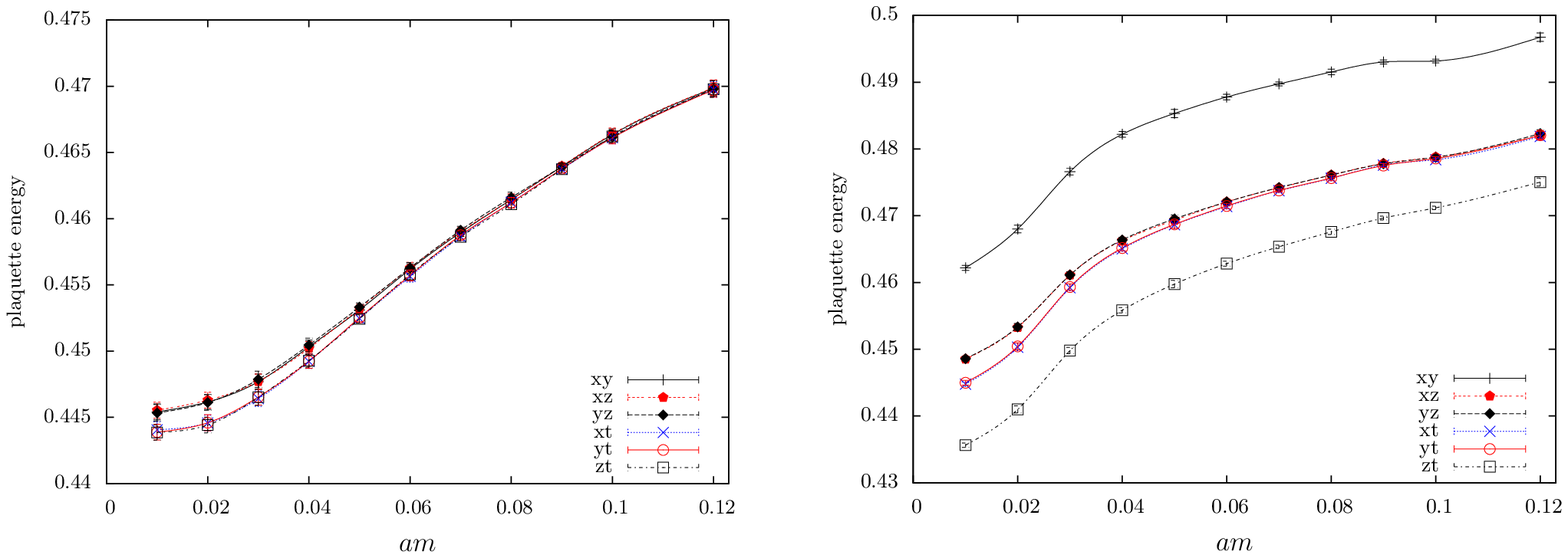}
\vspace*{-15cm}
\caption{Same as in \Fig{fig:fig10} but for $\beta=1.90$ (transition 
region).}
\label{fig:fig11}
\end{figure*}

\begin{figure*}[tb]
\includegraphics[width=1.0\textwidth]{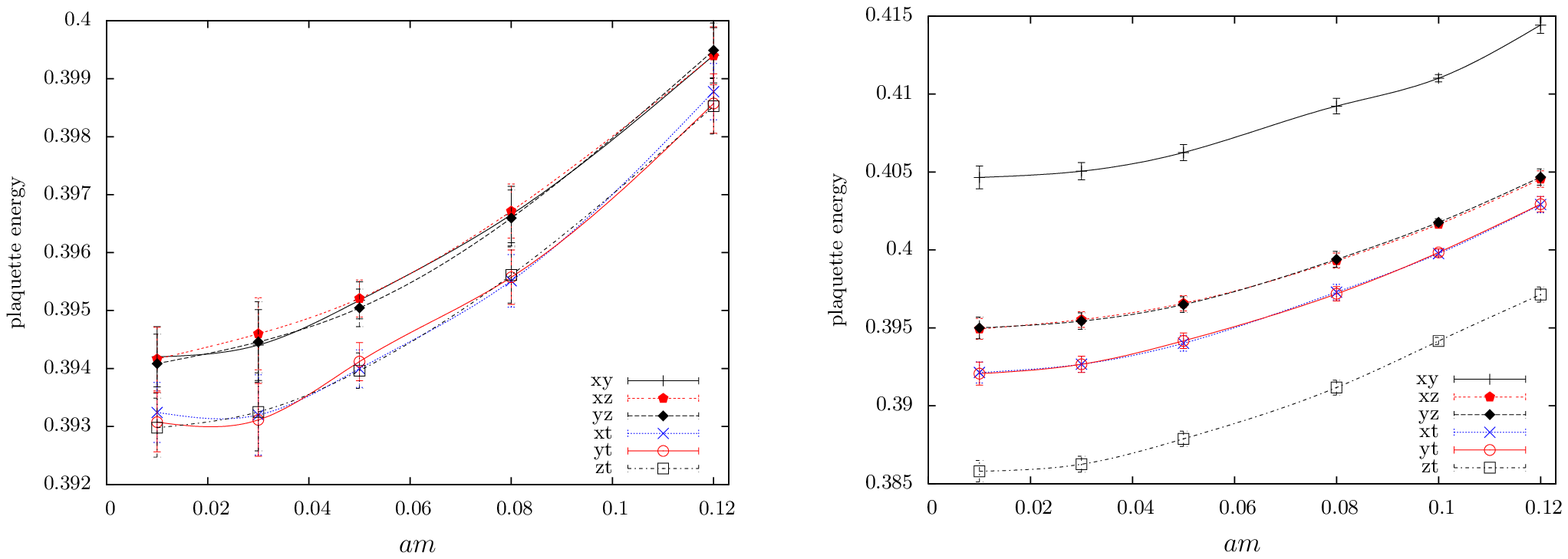}
\vspace*{-15cm}
\caption{Same as in \Fig{fig:fig10} but for $\beta=2.10$ (deconfinement 
phase).}
\label{fig:fig12}
\end{figure*}

In \Fig{fig:fig11} the same plots are shown for $\beta=1.90$ (in the transition 
region).  For vanishing magnetic field now the spacelike and the timelike 
plaquettes are different from each other for sufficient small quark mass. This is 
the well-known temperature effect observed in the deconfined phase and providing 
a non-vanishing energy density (see e.g \cite{Gavai:2005da}). However, in the 
right panel one can see that, with a strong magnetic field applied, the splitting 
of the $xy$ and the $zt$ plaquettes from the others is even stronger than the 
splitting between the other spacelike and timelike plaquettes.  Deep in the 
deconfined phase these results become even more pronounced, see \Fig{fig:fig12}.

\section{Conclusions}
\label{sec:conclusions}

We have investigated two-color QCD at finite temperature in an external 
magnetic field using lattice simulations. We have in particular studied 
how the magnetic field influences the chiral properties of the theory. 
As the chiral properties of this theory are similar to those of QCD, our
results should be relevant also for the physical case.

We have found for all temperatures for fixed bare quark mass that
the chiral condensate grows with the magnetic field.
Hence, the temperature of the chiral phase transition grows with the 
strength of the magnetic field. This is confirmed by a measurement 
locating the peak of the chiral susceptibility.
At the phase transition the pion mass is around 1.5 times the critical
temperature. A similar result had been found 
by \cite{D'Elia:2010nq,D'Elia:2011zu} in QCD with about the same ratio 
of the pion mass to the critical temperature. The opposite conclusion 
was reached in \cite{Bali:2011qj}, 
claiming that the transition temperature decreases with the magnetic field 
in QCD with the physical pion mass. It would be interesting to see if 
the result in our model is valid in the chiral limit, but we leave 
this to a future investigation. Furthermore, we have shown, by measuring the 
susceptibility of the Polyakov loop that the deconfinement and the chiral 
transitions move together to the same temperature, if small or  
large magnetic fields are switched on.

In this work, by extending our measurements to several values of the quark 
mass, we make a first investigation of the chiral limit of the theory
at three values of the temperature, one in the confined region, one in the 
transition region and another deep in the deconfined phase. In the chiral 
limit there is no additive renormalization necessary for the chiral condensate.
In the confined region, we find that the chiral condensate in the chiral 
limit seems to grow linearly with the magnetic field strength. This is in 
agreement with the prediction of the chiral model at zero temperature. 

At the other two values of the temperature chosen, the chiral condensate 
extrapolates to zero for all values of the magnetic field. Thus, in the
chiral limit there is a real chiral phase transition, 
which does not disappear for strong magnetic fields. 

We have also investigated the influence of the magnetic field on the gluonic
part of the theory. We find an asymmetry of the non-Abelian plaquette 
values with respect to the magnetic field. These variables are related to 
the gluonic part of the energy density of the theory. Thus we expect an 
influence of the magnetic field also on the equation of state. 
It would be interesting to conduct a study dedicated to this effect.

\section*{Acknowledgments}
E.-M.I. and M.M.-P. thank Mikhail Polikarpov who was the first to draw their 
attention to the importance of studying the behavior of non-Abelian gauge 
fields under the influence of external electromagnetic fields. 
We thank Edwin Laermann for providing us with the HMC code. 
A.S. gratefully acknowledges extensive help by Florian Burger in improving 
and translating the code to the parallel programming and computing platform 
CUDA.  We acknowledge useful discussions with Maria-Paola Lombardo,  
Falk Bruckmann, Rajiv Gavai and Sourendu Gupta.  

\bibliographystyle{apsrev}



\end{document}